# Organic Electronics Picks Up the Pace: Mask-Less, Solution Processed Organic Transistors Operating at 160 MHz

*Andrea Perinot, Michele Giorgio, Virgilio Mattoli, Dario Natali, Mario Caironi*[*]

Dr. A. Perinot, Dr. M. Giorgio, Prof. D. Natali, Dr. M. Caironi
Center for Nano Science and Technology@PoliMi, Istituto Italiano di Tecnologia, 20133 Milan, Italy
E-mail: mario.caironi@iit.it

Dr. Virgilio Mattoli
Center for Micro-BioRobotics, Istituto Italiano di Tecnologia, 56025 Pontedera (PI), Italy

Prof. Dario Natali
Department of Electronics, Information and Bioengineering, Politecnico di Milano, 20133 Milan, Italy

**Abstract**

Organic printed electronics has proven its potential as an essential enabler for applications related to healthcare, entertainment, energy and distributed intelligent objects. The possibility of exploiting solution-based and direct-writing production schemes further boosts the benefits offered by such technology, facilitating the implementation of cheap, conformable, bio-compatible electronic applications. The result shown in this work challenges the widespread assumption that such class of electronic devices is relegated to low-frequency operation, owing to the limited charge mobility of the materials and to the low spatial resolution achievable with conventional printing techniques. Here, it is shown that solution-processed and direct-written organic field-effect transistors can be carefully designed and fabricated so to achieve a maximum transition frequency of 160 MHz, unlocking an operational range that was not available before for organics. Such range was believed to be only accessible with more performing classes of semiconductor materials and/or more expensive fabrication schemes. The present achievement opens a route for cost- and energy-efficient manufacturability of flexible and conformable electronics with wireless-communication capabilities.

**Introduction**

The development of new applications in the fields of healthcare, energy, distributed sensing and entertainment will require the integration of electronic functionalities into everyday objects. Organic electronics has gained its place among the promising technologies to this purpose, owing to a set of distinctive features:[1] first, it is compatible with flexible substrates, which allows its integration with objects characterized by non-conventional form factors; second, it enables the use of deposition techniques derived from the graphic arts and gives access to cost-efficient manufacturing; third, selected organic materials are biocompatible, allowing for a high degree of integration between electronics and biology. The impressive progress in this field has been driven by: (i) the enhancement of a set of figures of merit, primarily the charge mobility of the semiconductors, now well exceeding amorphous silicon and rivalling low temperature deposited metal oxides;[2] (ii) the strengthening of cost- and energy-efficient fabrication strategies, with the notable examples of printing[3, 4] and laser processing,[5-7] which are now suitable for the micron-scale patterning of functional materials on large area; (iii) the demonstration of a set of proof-of-concept applications, including green/biodegradable electronic devices,[8] electronic skins and conformable patches for personal healthcare[9-12] or flexible organic microprocessors.[13]

However, in order to widen the set of applications that can be envisioned, a set of functionalities is still lacking. Among these, wireless communication between distributed electronic sensors/actuators and data-processing devices, or fast addressing capabilities for large-area arrays of sensors or light-emitting devices. The implementation of these functionalities would enable flexible large-area displays or sensor arrays and the creation of distributed wireless networks of electronic devices within the Internet of Things (IoT) framework.[14] So far, this set of applications has been considered out of reach for organic electronics.

A fundamental requirement to this goal is the realization of organic transistors, the basic building block of electronic circuits, operating at frequencies well above several tens of MHz and above. Such performance should also be obtained with the sole use of mask-less and scalable fabrication processes, in order to retain the manufacturability edge of organic devices.[15]

One of the most widely adopted figures of merit to quantify the maximum operation frequency of single transistors and allow comparison among different technologies is the *transition frequency $f_t$*, namely the frequency for which the ratio between the small-signal drain and gate currents is unity.[16] To date, the highest $f_t$ obtained for a an Organic Field-Effect Transistor (OFET) is 27.7 MHz .[17] Since $f_t$ is proportional to the bias voltage, some authors have used the *voltage-normalized transition frequency $f_t/V$* as a more convenient figure of merit to assess the relative performance of transistor technologies.[5, 18, 19] In this case, the highest $f_t/V$ value achieved for OFETs is 2.23 MHz/V,[20] achieved by virtue of a metal-oxide/self-assembled monolayer dielectric layer with high areal capacitance (700 nF/cm$^2$), a sub-micron channel length defined via high-resolution silicon stencil masks and extremely low contact resistance (29 Ωcm) between gold electrodes and a small-molecule organic semiconductor. These results however, together with the wide majority of the works on high-frequency OFETs, included masks and/or evaporation steps in the process flow.[21-24] Such an approach, while allowing the access to improved performances by virtue of an enhanced control over the deposition of the functional layers, poses a number of difficulties in terms of the future scalability to cost-efficient mass production. The sole use of mask-less direct-writing or solution-based techniques largely complicates the achievement of high-frequency operation, an issue also testified by the very limited number of attempts in the past.[5, 25-28]

As of now, despite the technologies and materials exhibiting the performances required for high-frequency operation in excess of several MHz and approaching the 100-MHz range (*i.e.* charge mobility approaching 1 cm$^2$/Vs and patterning resolutions below 1 µm) are in principle available, further progress has been hampered by a set of critical aspects that have been often overlooked. Primarily, the achievement of high effective charge mobility in downscaled transistors requires to obtain *normalized contact resistances* ($R_cW$) below 1 kΩcm (or less, depending on the other physical parameters and bias point of the transistor), which have been rarely demonstrated.[19] This aspect is intertwined with the need for reduction of the capacitive parasitism related to the gate-to-source and gate-to-drain geometrical overlap, which, in the frame of the current-crowding injection model, also affects charge-injection in a non-trivial way.[15] Finally, the design of efficient strategies for the dissipation of the generated heat becomes of paramount importance in order to prevent the destructive breakdown of the device and to allow for continuous-mode operation: downscaled OFETs with channel lengths in the order of the µm, sustaining a current per unit width in excess of 1 mA/mm and voltages in the range of few tens of volts, need to dissipate efficiently a power density in the range 10 to 100 Wmm$^{-2}$, which can easily lead to thermal breakdown of the device. The latter is not surprising, considering that the constituting materials, in particular plastic substrates, are characterized by a very low thermal conductivity, making heat dissipation highly inefficient. Recently, it was proposed that, for some applications (*e.g.* switching power converters, pulsed-mode data transfer), this can be circumvented by operating the transistor in pulsed mode, which allowed to reach a record $f_t$ of 40 MHz at a bias of 8.6 V in such operation regime.[29] However, fully exploiting the possibilities offered by a high-frequency organic technology requires continuous-mode operation, which in turn requires the adoption of efficient dissipation strategies.

Here we show that a route for the realization of high-frequency OFETs operating at a record-high $f_t$ of 160 MHz and $f_t/V$ of 4 MHz V$^{-1}$ can be implemented with a combination of scalable laser-based direct-writing techniques and solution-based deposition of organic polymers. We carefully selected a set of solutions to the problems illustrated above that complies with the requirement of a fully mask-less and solution-based process flow: these include laser-based patterning of metallic inks with a micron-scale resolution, the modification of the electrodes with a self-assembled monolayer for the achievement of low contact resistance and the adoption of a substrate with high thermal conductivity. With this result, we prove that operational frequencies in excess of 100 MHz can be achieved with organic transistors. Moreover, we do not only show a working organic transistor with the highest $f_t$ to date and the highest $f_t/V$ for continuous operation, but we also demonstrate that a route for the achievement of this performance with scalable, mask-less, solution-based techniques is available, and that the future implementation of cost- and energy-efficient mass manufacturing of high-performance organic electronic applications is credible.

**Results**

We realized high-frequency OFETs in a bottom-contact, top-gate architecture with the layout schematized in **Figure 1a**, carefully selecting the architecture, materials and processes in order to overcome a variety of limitations to high-frequency operation.

Our fabrication process relies on the flow illustrated in Figure 1b. We selected femtosecond-laser sintering as a direct-writing patterning technique for the realization of micron-scale conductive electrodes for OFETs. Such an approach was successfully adopted in the past for the realization of metallic grids[30] and OFETs,[7, 31, 32] including high-frequency, direct-written and printed OFETs,[27, 33] also on plastic substrate.[5] The choice of the proposed fabrication scheme is advantageous for a variety of future implementations into a wide set of applications, by virtue of its digital nature and compatibility with different substrate materials. However, devices of the kind we realize in this work, when fabricated on plastic, are prone to suffer of thermal runaway or breakdown (described later in the text), due to the significant amount of power density in the channel region of the device and to the limited thermal dissipation properties of plastics (commonly exhibiting thermal conductivity in the range 0.1 – 0.5 $Wm^{-1}K^{-1}$). To comply with the need for efficient thermal dissipation of such generated heat, we adopted here a highly thermally-conductive substrate of aluminum nitride (AlN), exhibiting a thermal conductivity in the order of 170 $Wm^{-1}K^{-1}$.

To fabricate our OFETs we first coat our substrate with an Ag-nanoparticle ink, then we locally induce the agglomeration of the metal nanoparticles into conductive structures via laser sintering.[5] Then, the unprocessed part of the ink is washed out with an organic solvent, leaving high-resolution conductive patterns with a thickness of 70 nm on the substrate. These structures will constitute the source and drain electrodes of the realized OFETs, yielding a channel length $L$ = 1.2 µm, a channel width $W$ = 800 µm, an electrode width $L_c$ = 1.7 µm. To promote an efficient charge injection from such electrodes into the semiconductor, we then induce the self-assembly of a monolayer of dimethylamino(benzenethiol) (DABT) on the surface of the metallic patterns.[34] Then, we adopt the widely-studied and good electron transporting semiconducting co-polymer poly[N,N'-bis(2-octyldodecyl)-naphthalene-1,4,5,8-bis(dicarboximide)-2,6-diyl]-alt-5,5'-(2,2'-bithiophene), P(NDI2OD-T2), and deposit a thin layer of such material via off-centered spin-coating from a solution in toluene. Such a selection of deposition technique and solvent yields a semiconducting layer with enhanced charge transport properties thanks to the promotion of aggregates formation in the solution, which in turn yields the formation of a layer of aligned polymer nanofibrils.[35, 36] We then adopted a bilayer dielectric: we first deposit a 40-nm-thick layer of polystyrene blended with an azide-based crosslinker (1,11-Diazido-3,6,9-trioxaundecane) and we cross-link such layer via UV-light exposure at a wavelength of 256 nm. On top of the polystyrene interlayer, we spin-coat a 300-nm-thick layer of poly(vinyl cinnamate), which is then analogously photo-crosslinked. The complete dielectric bilayer exhibits an areal capacitance $C_{diel}$ = 8.54 nF cm$^{-2}$, calculated using the literature value of 3.4 for the dielectric constant of poly(vinyl cinnamate) and a value of 2.6 for cross-linked polystyrene (determined from our measurements on capacitor devices). The top gate electrode is then realized via laser sintering in correspondence of the transistor channel, keeping the overlap with source and drain electrodes low, to comply with the need of reducing the overlap capacitive parasitism. This is the first time laser sintering[31] is used for the fabrication of gate electrodes on polymer dielectrics in top-gate structures. Encapsulation of the device to prevent degradation induced by the exposure to the ambient environment concludes the fabrication; further details are reported in the Supporting Information.

A top-view representation of the final device is shown in Figure 1c alongside with a magnified micrograph of the active region of the transistor, which highlights the fine alignment between the top gate electrode and the channel area. We confirmed such alignment, associated with a low capacitive parasitism, with cross-sectional SEM imaging of the device (Figure 1c), which allows to estimate the size of the geometrical overlap between electrodes in the range ∼ 0-250 nm (Figure S1).

We measured the DC transfer (**Figure 2a**) and output characteristics (Figure S3) of our transistors, verifying a correct operation up to a bias voltage of 40 V, with a maximum gate leakage current in the order of the nA, with respect to a channel current in the order of few mA. This proves that laser processing on top of a multilayer stack of organic materials, including a semiconductor and a dielectric, is compatible with the fine patterning of high-resolution conductive electrodes without damage to the underlying materials.

We then highlight how the integration of a substrate with a high thermal conductivity in our process allows ideal DC operation of the device and prevents the thermal breakdown. In particular, in the case of OFETs with the same architecture and comparable fabrication process, realized on a glass substrate (which exhibits a lower thermal conductivity in the order of 1 $Wm^{-1}K^{-1}$), when the generated power per unit area approaches the range 20-30 W $mm^{-2}$, the devices start to suffer from thermal degradation, the current driven by the device saturates/drops with respect to the increase of the gate voltage and severe hysteresis appears in the transfer curve (Figure S4). Contrarily, for the devices of this work, even at the bias point corresponding to the maximum generated power per unit area $P_{th}$ ($I_d$ = 2.18 mA, $V_d$ = 40 V and $P_{th}$ = 90 W $mm^{-2}$), correct operation of the device is preserved and no signs of thermal degradation are visible.

We calculated the apparent charge mobility of our devices in the linear ($\mu_{lin}$) and saturation regimes ($\mu_{sat}$) versus gate voltage (Figure 2b). The ideality of the DC operation of the transistors is confirmed by the flatness of the curves in the fully accumulated regime above 10 V, with a slight roll-off that can be attributed to some residual impact of the contact resistance. The maximum values for the apparent charge mobility are $\mu_{lin}$ = 0.22 cm$^2$V$^{-1}$s$^{-1}$ and $\mu_{sat}$ = 0.62 cm$^2$V$^{-1}$s$^{-1}$. We extracted the width-normalized contact resistance $R_cW$ and the intrinsic charge mobility $\mu_i$ of our devices, which we estimate to be $R_cW$ = 300 Ωcm and $\mu_i$ = 1 cm$^2$V$^{-1}$s$^{-1}$ in the saturation regime at a bias point of $V_g = V_d$ = 40 V (see Supplementary Information for details). Such a value of $R_c$ is not only a key requirement in order to access frequency regimes in excess of 100 MHz,[15, 19] but is among the best reported values for OFETs in general and is extremely low when considering the case of transistors realized via direct-writing, solution-based methods and optimized for low geometrical overlap of electrodes and high frequency operation.[37-39]

We then measured the AC characteristics of our device by means of S-parameters, using a setup already described in our previous work,[27] calibrated with a SOLT procedure and corrected with a 12-term error model. From the measured S-parameters, the parasitic contributions of the pads and interconnections are de-embedded from the measurement with a one-step procedure[40] and the hybrid parameter $h_{21}$ is extracted (Figure 2c), allowing to identify $f_t$ according to $h_{21}(f_t)$ = 0 dB, which yields an unprecedented $f_t$ of 160 MHz at a bias voltage of 40 V for OFETs in the case of the best device (Figure 2c, linear fit). In terms of the voltage-normalized transition frequency $f_t/V$, we reached a figure as high as 4 MHz V$^{-1}$, also in this case the highest value reported for an OFET. Such extracted $f_t$ performance is robust with respect to thermal degradation effects: measurements on a nominally identical device results in a practically identical $f_t$ of 158 MHz, which remains stable after a second, consecutive measurement of $h_{21}$ at $V_g$ = 40 V (Figure 2d, black and red lines) and after further measurements at gate biases of 35 and 30 V (Figure 2d, blue and purple lines).

As a crosscheck of the consistency of the AC performance, we extracted the values for the gate/drain and gate/source capacitances $C_{gd}$ and $C_{gs}$ for $V_g = V_d$ = 40 V, alongside with the total gate capacitance $C_g = C_{gd} + C_{gs}$ (Figure S5). The total gate capacitance, at a first order, can be estimated as follows:

$$C_g \cong C_{diel} W \left( \frac{2}{3} L + 2 L_{ov} + 2d \right) \quad (1)$$

where $L_{ov}$ is the geometrical overlap between gate and source (or drain) electrode and $2d$, for low-overlap structures of the kind presented here, accounts for the contribution of the fringing field in the form of an "equivalent overlap length", equal to the thickness of the dielectric $d$.[15] According to this formula, and with $L_{ov}$ in the range 0-250 nm, the total gate capacitance $C_g$ can be estimated to be in the range 101 - 135 fF, which is in good agreement with the value extracted from our measurement (140 – 150 fF above 30 MHz, Figure S5). The transconductance and output resistance can be estimated from the DC curves respectively as

$g_m = \frac{dI_d}{dV_g}$ and $r_o = \left(\frac{dI_d}{dV_d}\right)^{-1}$, evaluated at the transistor bias $V_g = V_d = 40$ V, yielding $g_m$ = 0.115 mS and $r_o$ = 25.3 kΩ. These values obtained from the DC characterization are in agreement with the S-parameters measurements ($g_m$ = 0.115 mS and $r_o$ = 22 kΩ at 10 MHz, Figure S6). In addition, we verified that $g_m$ is not altered by the de-embedding procedure, confirming the consistency of the obtained results (Figure S6).

The measured $f_t$ can be compared to the theoretical value estimated from the transistor DC electrical parameters and geometrical dimensions, according to:

$$f_t = \frac{g_m}{2\pi C_g}. \quad (2)$$

With the range of values for $C_g$ calculated above and with the range of values for $g_m$ extracted from DC, the theoretical $f_t$ is calculated to be in the range ~ 140 - 180 MHz, which is consistent with our measured value. By including our additional analysis on the contact resistance (see Supplementary Information), the measured $f_t$ can also be related to the value predicted by more refined theoretical models in recent reports,[15, 19] which include not only the effects of the fringing electric field for low-overlap structures (already accounted for by Equation (1)) but also the effects associated with charge injection physics in staggered OFETs with small electrode overlap. The application of such model consistently returns, for the parameters of the transistors of this work, a predicted $f_t$ in the range 138-146 MHz (see Supplementary Information), which is not dissimilar to our measured result.

Overall, high-frequency operation at 160 MHz of solution-processed OFETs is demonstrated via an S-parameter measurement and further validated by the agreement of the extracted transistor small-signal AC parameters with the ones calculated through physical and geometrical considerations. This experimental demonstration agrees with and complements the theoretical roadmaps described in recent works.[15, 19]

**Discussion**

Contrarily to the widespread assumption that organic electronics is relegated to very low-frequency operation, we have shown here that organic FETs can operate at an $f_t$ of 160 MHz and $f_t/V$ of 4 MHz V$^{-1}$. This value of $f_t$ is by far the highest reported for any organic transistor to date, while $f_t/V$ is the best reported for organic transistors capable of sustaining continuous biasing (Table S1). The significance of this achievement is further reinforced by the sole adoption of direct-writing and solution-based fabrication methods, which have traditionally complicated the achievement of high-performance figures of merit, as well as finely-controlled patterning of functional materials at the micron scale.

The OFET AC performance demonstrated in this work was achieved both by devising a set of strategies to overcome the bottlenecks to high-frequency operation and by combining them into a fabrication scheme solely using scalable techniques. First, the high patterning resolution necessary both to downscale the transistor dimensions and to contain the capacitive parasitism has been achieved by using laser sintering, which allowed the fine alignment of micron-sized electrodes via direct writing. Second, the charge injection from the contacts, which must be very efficient for downscaled architectures with low overlap between gate and bottom electrodes, has been promoted by inducing the self-assembly of an amine-based monolayer. This approach allowed to achieve width-normalized contact resistance $R_cW = 300$ Ωcm, which is among the best reported values for solution-processed, direct-written OFETs in general. This achievement is further reinforced by the fact that it is associated with an architecture optimized for high-frequency operation, whose low electrode overlap is well-known to be detrimental for charge injection. Third, thermal breakdown/degradation has been avoided by using an appropriate thermally-conductive substrate. The latter result highlights an unprecedented need for substrate materials for OFETs, combining flexibility and sufficient thermal conductivity, thus indicating a clear path to be further pursued in future.[41-44]

In conclusion, we have demonstrated that high-frequency operation in excess of 100 MHz is accessible to organic-based electronics. The result we show here represents a suitable complement and a validation to a set of recent reports that theoretically detailed a feasible roadmap towards high-frequency operation or organic transistors.[15, 19] Within the roadmap detailed in such works, our achievement of a $R_cW$ of 300 Ωcm in high-frequency devices based on printed polymers constitutes one of the key enablers.

These achievements challenge the conventional, well-known tradeoff between the higher electrical performances of inorganic materials (*e.g.* silicon, metal-oxides, carbon nanotubes) with the advantageous mechanical properties and the cost- and energy-efficient processability of organics. Our findings, overall, outline a credible route towards the adoption of organics in an expanded set of applications, including remote healthcare, distributed sensing, design and entertainment, requiring the availability of a technology integrating large-area electronics with wireless-communication capabilities, realized via cost- and energy-efficient production schemes.

**Experimental Section**
For the experimental section, please refer to the Supporting Information.

**Acknowledgements**
The authors are grateful to L. Criante for the support with the femtosecond laser machining setup. Part of the work has been carried out at Polifab, the micro- and nanotechnology center of the Politecnico di Milano. This work was financially supported by the European Research Council (ERC) under the European union's Horizon 2020 research and innovation programme "HEROIC", grant agreement 638059.

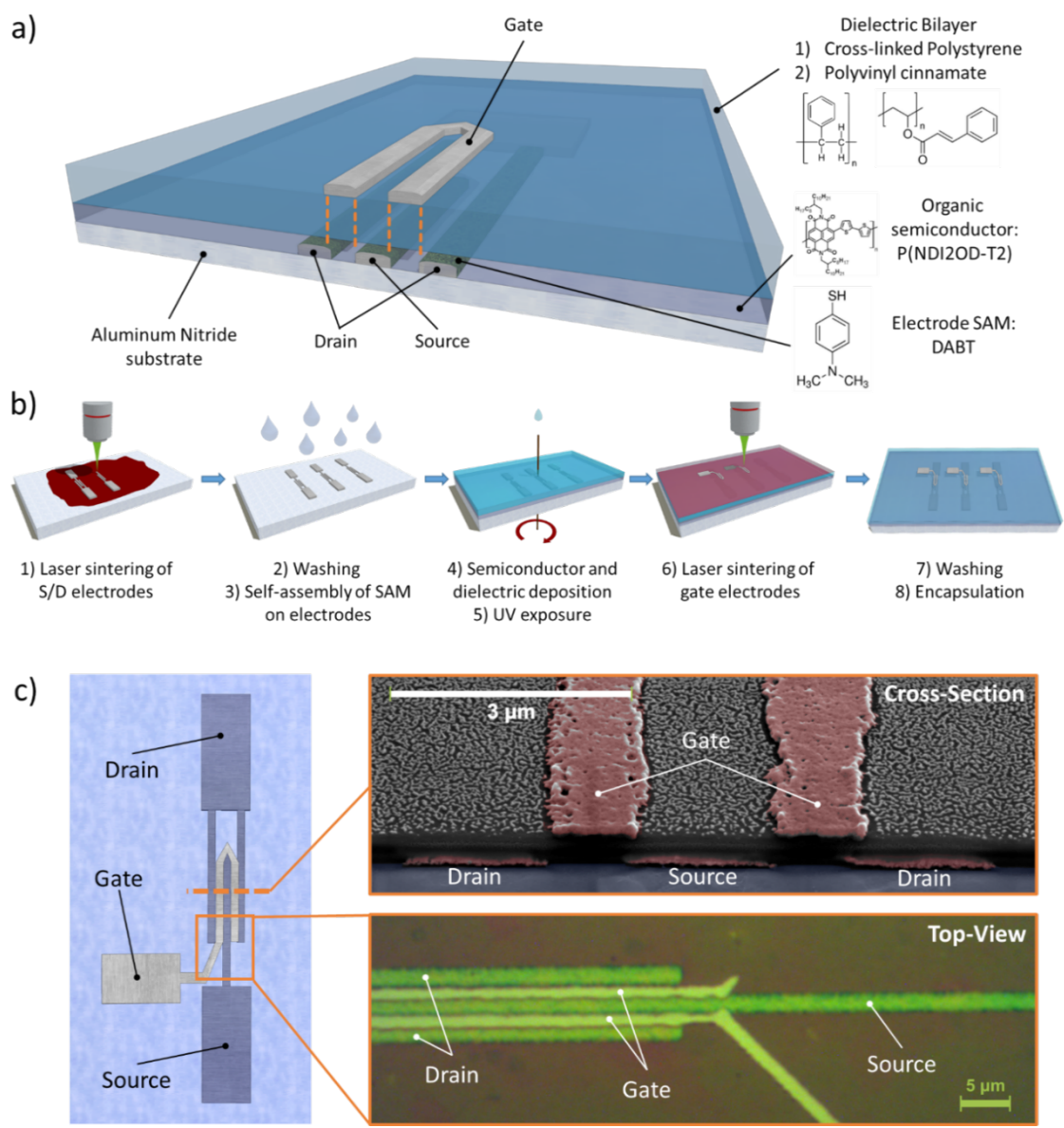

**Figure 1:** a) 3D sketch of the realized OFET architecture (not in scale), alongside with the adopted materials. b) Scheme of the device fabrication process. c) Sketch of the realized device (not in scale), with a top-view micrograph of the active region (bottom right, rotated by 90°) and a cross-sectional SEM image of the active area and the electrodes (top right). The particulate residuals on the top layer, at the sides of the gate electrode, are part of a gold coating specifically deposited for the imaging, not present in the measured devices.

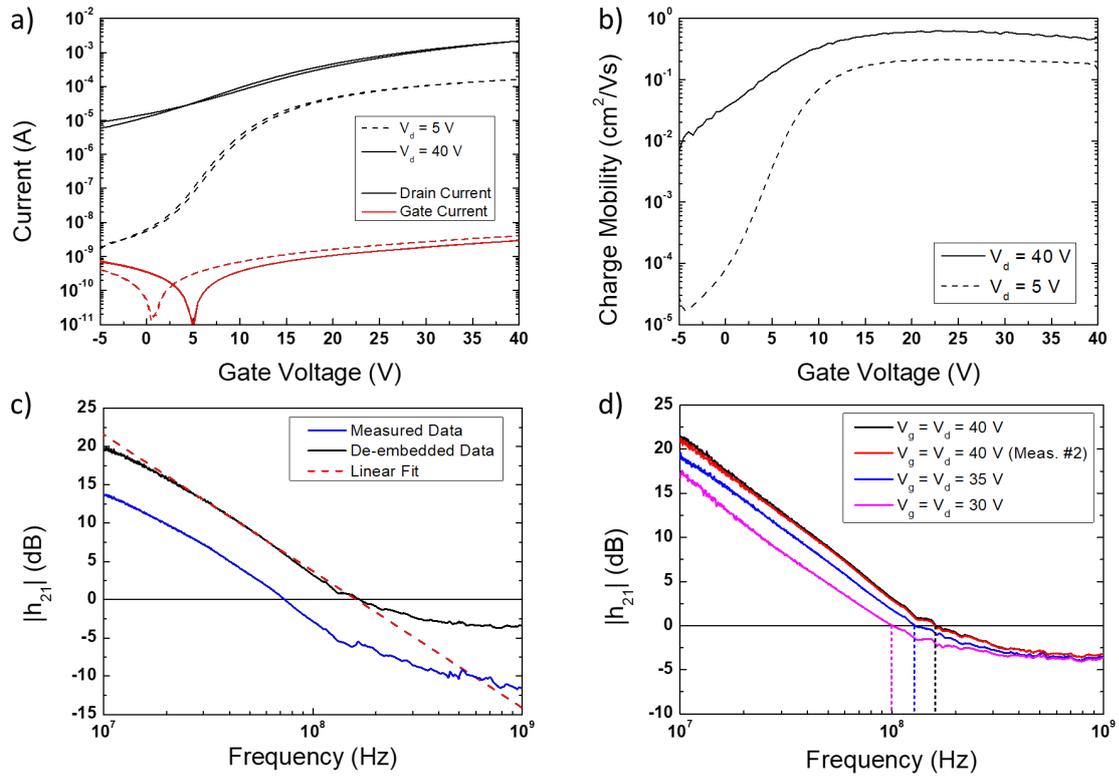

**Figure 2:** a) Measured transfer curve and b) apparent charge mobility for our high-frequency OFET. c) $h_{21}$ for the same device, extracted from the S-parameter measurement at $V_g = V_d =$ 40 V. d) Consecutive measurements of $h_{21}$ for another, nominally identical device, at different bias voltages in the range 30 to 40 V (de-embedded data). No appreciable degradation effects are visible after two consecutive measurements at a bias of 40 V.

# Supporting Information

**Organic Electronics Picks Up the Pace: Mask-Less, Solution Processed Organic Transistors Operating at 160 MHz**

*Andrea Perinot, Michele Giorgio, Virgilio Mattoli, Dario Natali, Mario Caironi\**

**Methods**

*Materials:* Aluminum Nitride substrates were purchased from MARUWA CO. The Ag-nanoparticle ink used in laser sintering (NPS-L) was purchased from HARIMA. Dimethylamino(benzenethiol) was purchased from TCI Chemicals. P(NDI2OD-T2) was purchased from Polyera. Polystyrene ($M_w$ = 2000000), poly(vinyl cinnamate) ($M_n$ = 40000) and 1,11-Diazido-3,6,9-trioxaundecane were purchased from Sigma Aldrich. The epoxy resin was purchased from Robnor.

*FET fabrication:* On the AlN substrates we first defined, via conventional photolithography, a set of structures and calibration patterns for connecting our FETs to the high-frequency probes for S-parameter measurement. The details on the structures can be found in *Giorgio et al.*[1] Then, we coated the Ag-nanoparticle ink onto these substrates via spin-coating at 7000 rpm for 5 min. Then, we patterned the source and drain bottom electrodes through laser sintering using the setup and following the procedures illustrated in our previous work.[2] In this case, the incident laser power was 17.2 mW at a scanning speed of 0.05 mm s$^{-1}$. The unprocessed part of the ink was removed by thorough rinsing with o-xylene. Then, Ar-plasma is applied for 4 minutes at a power of 100 W, and the self-assembly of DABT on the silver electrodes is induced by dipping the samples in a solution of 17 µl DABT in 12 ml of isopropanol for 15 minutes. The samples are then rinsed with isopropanol. The semiconductor layer is then deposited via off-centered spin-coating[3] (in nitrogen atmosphere) of a 7 g/l solution of P(NDI2OD-T2) in toluene, at a speed of 1000 rpm for 30 s. The samples are then annealed at 100 °C for 15 minutes. After cooling, a 40-nm-thick layer of polystyrene, mixed with 1,11-Diazido-3,6,9-trioxaundecane at a weight ratio of 10:1, is deposited via spin-coating at a speed of 1500 rpm for 5 minutes from a solution in n-butyl acetate at a concentration of 7.5 g/l. Then, we spin-coated a solution of 50 g/l poly(vinyl cinnamate) in cyclopentanone at a speed of 1500 rpm for 2 minutes, so to yield a 300-nm-thick layer, which is then cross-linked analogously to the underlying layer. We then patterned the gate electrodes via laser sintering with the same procedure as illustrated above, using an incident power in the range 4.9-5.3 mW and a scanning speed of 0.02 mm s$^{-1}$. Finally, we encapsulated the devices by spin-coating a 50 g/l solution of PMMA in o-xylene at a speed of 1300 rpm for 60 s, followed by annealing at 60 °C for 20 min for solvent removal, followed by deposition of a 1-um-thick layer of parylene via CVD, finally completed by drop-casting a bi-component epoxy resin. After 24 h, the samples are then annealed for 8 h in nitrogen atmosphere at 105 °C.

*Measurement:* The thickness of the laser-sintered electrodes and of the polymer layers were measured with an Alpha-Step IQ profilometer by KLA-Tencor. The DC measurements were performed in nitrogen atmosphere using a Keysight B1500A Semiconductor Parameter Analyzer. The AC measurement was performed in ambient atmosphere using a setup and calibration method already described previously.[1] The parasitism attributed to the measurement pads and interconnections has been removed by measuring an open structure with a geometry identical to the interconnections used for the transistor measurement.[1]

**Supporting Figures**

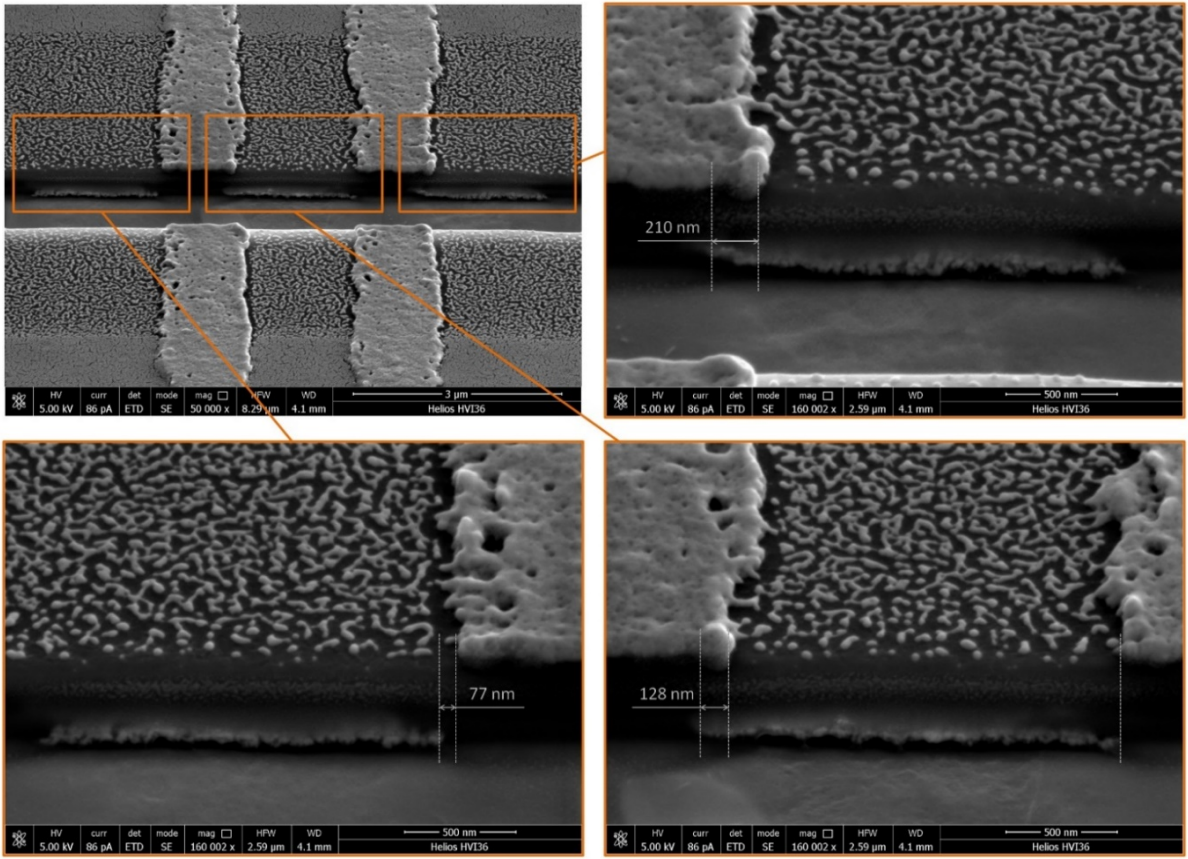

**Figure S1:** Cross-sectional SEM images of the realized device, with magnifications of the area in the vicinity of the bottom electrodes. Measurements of the electrode geometrical overlap are also shown.

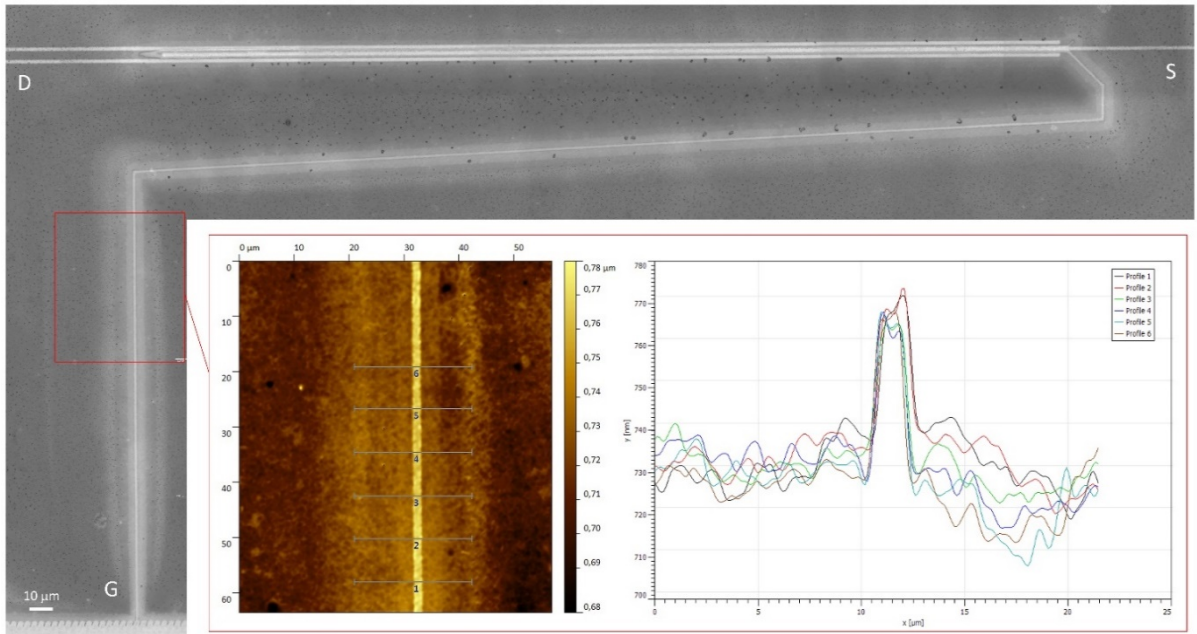

**Figure S2:** Optical image of a typical device and related confocal profilometry highlighting a particular of the laser-sintered gate track. The average thickness of the track on top of the dielectric stack is ~ 40-50 nm. All images were acquired with a Leica DCM 3D Confocal Profilometer, at 150x magnification. Profilometer data were elaborated with Gwyddion software (plane tilting, profile extraction, file conversion).

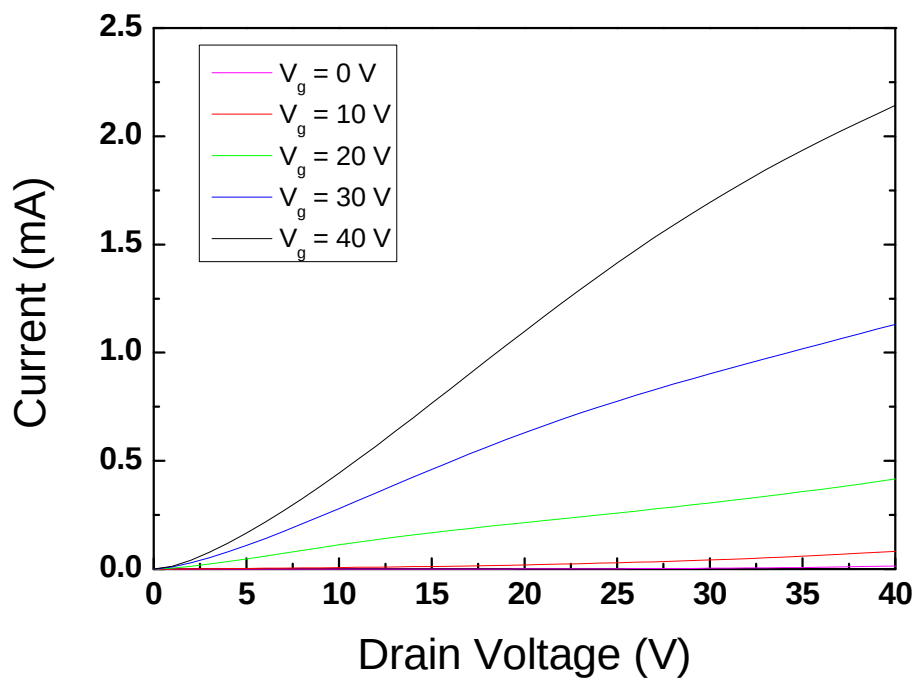

**Figure S3:** Measured output curve for the realized high-frequency OFET.

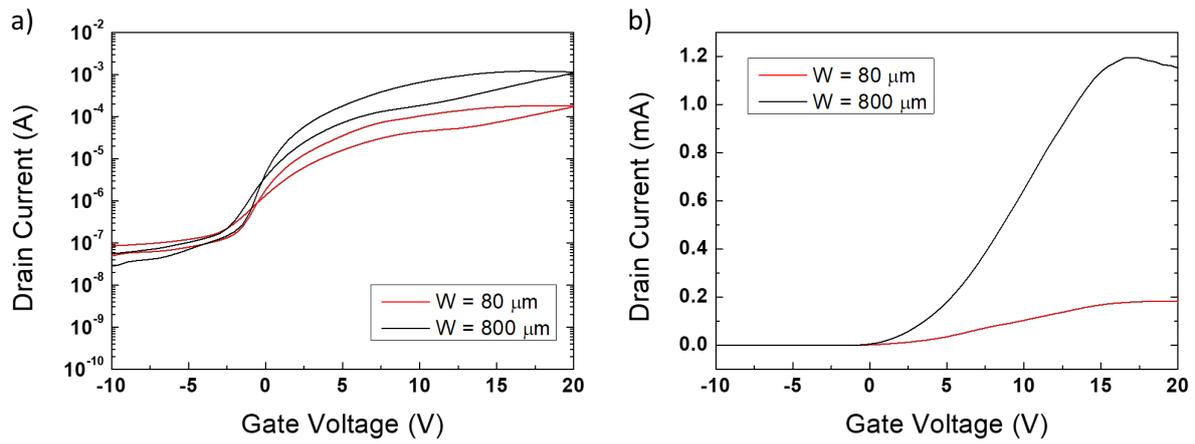

**Figure S4:** Measured transfer curves for OFETs on a glass substrate with low thermal conductivity, in the order of 1 W/mK. The devices are fabricated with the same architecture and comparable process as the ones fabricated on AlN substrate, and differ in terms of channel length and dielectric material (in this case, $L$ = 1.4 µm, poly(vinyl alcohol) is used in place of poly(vinyl cinnamate) and the channel width is $W$ = 800 µm or $W$ = 80 µm). a) Transfer curves for $V_d$ = 20 V in logarithmic scale. b) Same transfer curves (only forward scan) in the linear regime.

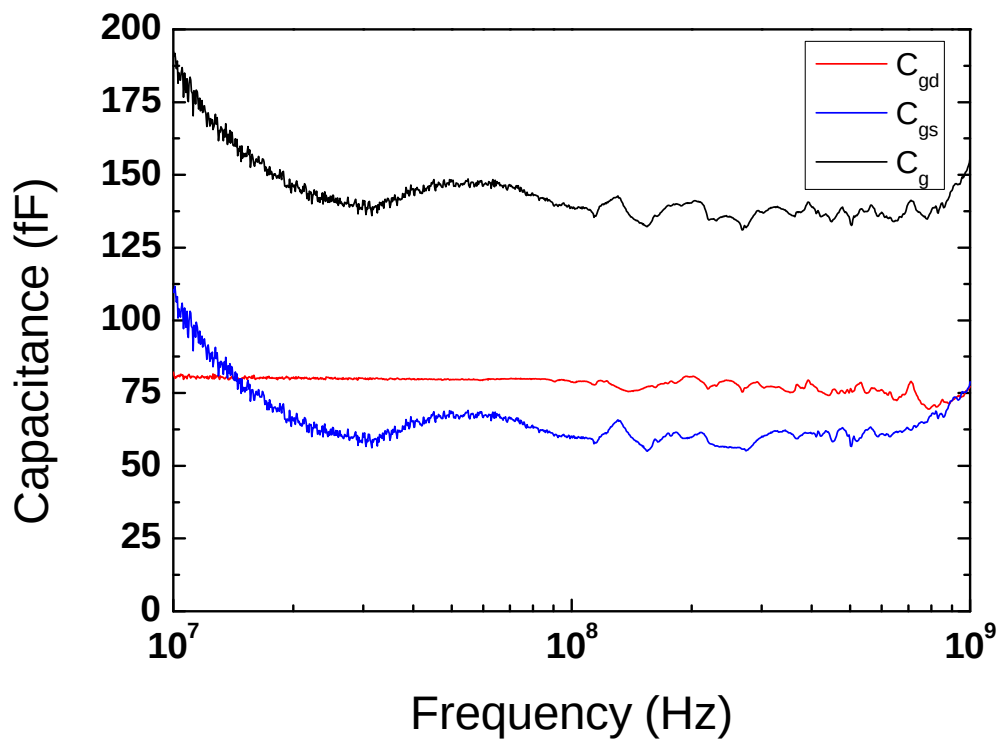

**Figure S5:** Calculated gate capacitances, extracted from the S-parameter measurement.

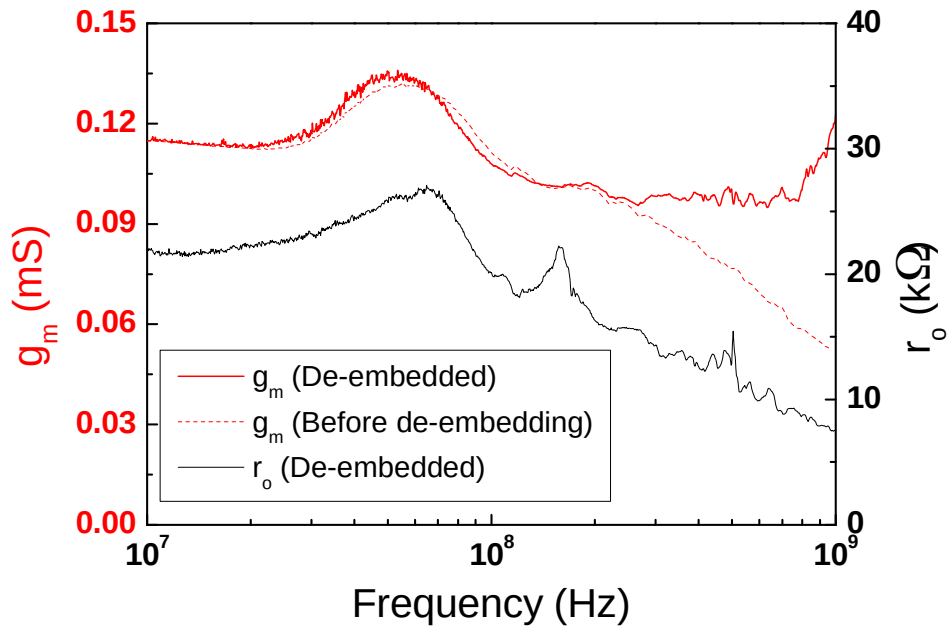

**Figure S6:** Calculated $g_m$ and $r_o$, extracted from the S-parameter measurement. We show the extracted value for $g_m$ both before and after the de-embedding.

**Table S1**: Selected results in the literature for high-frequency organic transistors and circuits, ordered in terms of $f_t/V$. Here reported only the works exhibiting $f_t/V$ in excess of 1 MHz/V, in continuous-mode operation.

| Reference | $f_T$ (MHz) | $f_t/V$ (MHz/V)[a] | Flexible Substrate | Mask-less fabrication |
|---|---|---|---|---|
| *This Work* | 160 | 4 | X | Yes |
| *Borchert et al.*[4] | 6.7 | 2.23 | Yes | X |
| *Perinot et al.*[2] | 14.4 | 2.06 | Yes | Yes |
| *Yamamura et al.*[5] | 20 | 2 | X | X |
| *Nakayama et al.*[6] | 19 | 1.9 | X | X |
| *Uno et al.*[7] | 25 | 1.67 | X | X |
| *Giorgio et al.*[1] | 19 | 1.27 | X | Yes |
| *Kitamura et al.*[8] | 27.7 | 1.11 | X | X |
| *Uemura et al.*[9] | 20 | 1 | X | X |

**a) Our calculation when not reported. $f_t$ is normalized to the highest voltage between source-gate or drain-gate. X: not applicable.**

**Extraction of the contact resistance**

In the saturation regime, which is the case of interest here, only the contact resistance at source side matters (provided that voltage drop on the contact resistance at drain side is low enough to maintain the transistor in saturation[10]).

In the framework of the current crowding model, suitable for staggered transistors, contact resistances can be expressed as:

$$R_C = \frac{R_y}{W L_0 \tanh\left(\frac{L_{ov}}{L_0}\right)}, \tag{1}$$

Where: $L_{ov}$ is the gate-contact overlap; $R_y$ is the resistance per unit area taking into account injection and transport across the bulk; $L_0 = \sqrt{R_y/R_{sh}}$ is the injection length, viz. the characteristic length over which injection would take place for very large $L_{ov}$, $R_{sh}$ being the channel sheet resistance. Modelling the carrier mobility as a power law, $\mu = \mu_0 (V_G - V_T)^\gamma$, the sheet resistance can be expressed as $R_{sh} = \left[\mu_0 C_{ins}(V_G - V_T)^{\gamma+1}\right]^{-1}$.

For the case of very small $L_{ov}$, which is the case of interest here (actually for $L_{ov} < L_0$, to be verified *a posteriori*), Equation (1) can be simplified as the sum of a constant term and of a $V_G$-dependent term, as it follows[11]:

$$R_C = \frac{R_y}{W L_{ov}} + \frac{1}{3}\frac{L_{ov}}{W} R_{sh} = R_{c,const} + R_{c,var}(V_G), \tag{2}$$

where the first term accounts for injection and transport across the film, whereas the second term accounts for transport along the film at the semiconductor/insulator interface.

The challenge in the saturation regime is due to the fact that the current voltage relationship incorporating the effects of contact resistance is actually an implicit function, without the possibility of writing current as an explicit function of $V_G$ in the general case:

$$I = \mu_0 C_{ins} \frac{W}{L}(V_G - V_T - R_C I)^{\gamma+2}, \tag{3}$$

where $V_T$ is the threshold voltage. There are 5 unknowns in Equation (3): $\mu_0$, $\gamma$, $V_T$, $R_{c,\,const}$, $L_{ov}$, ($R_{c,\,var}$ can be expressed as a function of $\mu_0$, $\gamma$, $V_T$, $L_{ov}$). To extract them from experimental data, we devise an iterative fitting algorithm. In addition, to ease the procedure and reduce the number of fitting parameters, we select reasonable ranges for $\mu_0$ and $V_T$, and for each ($\mu_0$, $V_T$) couple we run the following algorithm.

The parameter $\gamma$ is initialized at 0.01.

1. Since $\mu_0$ and $V_T$ are fixed and $\gamma$ is initialized (or fitted, *vide infra*), we can calculate $\widehat{V}_G$, the base which is raised to ($\gamma + 2$) in Equation (3):

$$\widehat{V}_G = V_G - V_T - R_C I = \left( \frac{I}{\mu_0 C_{ins} \frac{W}{L}} \right)^{\frac{1}{\gamma+2}}. \qquad (4)$$

2. Now we take advantage of the fact that: $V_G$ and $I$ are experimentally measured; $\mu_0$ and $V_T$ are fixed. We plot $V_G - V_T - R_C I$ versus $V_G$ and, exploiting Equation (2), we fit $R_{c,const}$, $L_{ov}$ and $\gamma$, with the constraint $\gamma > 0$. The fitting is done in the range 23 V $< V_G <$ 40 V.

3. With the value for $\gamma$ estimated at step 2, we jump to step 1 and reiterate for 100 cycles.

We sometimes experienced oscillations in the fitted value for $\gamma$ between 0 and a certain $\tilde{\gamma}$. Indeed, for consistent and realistic fitted parameters, $\tilde{\gamma}$ is very close to 0 (actually smaller than 0.043), therefore the impact of such oscillation is negligible. In these cases, to proceed with the analysis, we arbitrarily chose $\gamma = \frac{\tilde{\gamma}}{2}$ and we run a final direct fit of $R_c$, determining $R_{c,const}$ and $L_{OV}$. Later, we verified that different choices for $\gamma$ (*i.e.* $\gamma = \frac{\tilde{\gamma}}{4}$ or $\gamma = \frac{3}{4}\tilde{\gamma}$) did not appreciably change the results of the fitting.

The parameters $\gamma$, $R_{c,const}$ and $L_{ov}$ extracted with $\mu_0$ in the range 0.94 – 1.1 cm²/Vs and $V_T$ in the range 5.9-6.2 V are shown below in Table S2. From the independent measurement of the geometrical overlap between electrodes and of the dielectric thickness, within the framework of the gate capacitance model illustrated in the main text,[12] we identify the acceptable values for $L_{ov}$ (i.e. 0.34 µm < $L_{ov}$ < 0.61 µm) and we highlight the corresponding combinations in red in Table S2.

**Table S2:** Extracted values of γ, $R_{c,const}$, $L_{ov}$ as a result of the fitting of the experimental curves according to our algorithm. Values corresponding to the combinations where $L_{ov}$ is within the acceptable range (according to a second independent measurement) are highlighted in red.

a) $R_{c,const}$ (Ω)

| | Vt=5.9 V | Vt=5.92 V | Vt=5.94 V | Vt=5.96 V | Vt=5.98 V | Vt=6 V | Vt=6.02 V | Vt=6.04 V | Vt=6.06 V | Vt=6.08 V | Vt=6.1 V | Vt=6.12 V | Vt=6.14 V | Vt=6.16 V | Vt=6.18 V | Vt=6.2 V |
|---|---|---|---|---|---|---|---|---|---|---|---|---|---|---|---|---|
| mu0=0.94 | 7758 | 7952 | 8202 | 8540 | 9019 | NA | NA | NA | NA | NA | NA | NA | NA | NA | NA | NA |
| mu0=0.96 | 6010 | 5990 | 5966 | 5938 | 5904 | 5861 | NA | 5738 | 5641 | 5498 | 5316 | NA | NA | NA | NA | NA |
| mu0=0.98 | 5155 | 5086 | 5009 | 4920 | 4819 | 4702 | 4564 | 4399 | 4200 | 3952 | 3661 | 3248 | 2678 | 2629 | 2643 | 2657 |
| mu0=1 | 4654 | 4574 | 4486 | 4389 | 4280 | 4158 | 4021 | 3865 | 3686 | 3478 | 3250 | 2960 | 2636 | 2650 | 2663 | 2677 |
| mu0=1.02 | 4329 | 4248 | 4161 | 4066 | 3962 | 3848 | 3723 | 3584 | 3429 | 3256 | 3072 | 2850 | 2655 | 2669 | 2683 | 2697 |
| mu0=1.04 | 4102 | 4025 | 3941 | 3852 | 3755 | 3651 | 3537 | 3414 | 3279 | 3134 | 2978 | 2797 | 2674 | 2688 | 2702 | 2716 |
| mu0=1.06 | 3937 | 3864 | 3785 | 3701 | 3612 | 3516 | 3413 | 3303 | 3183 | 3059 | 2922 | 2770 | 2692 | 2707 | 2721 | 2735 |
| mu0=1.08 | 3813 | 3743 | 3669 | 3591 | 3508 | 3420 | 3326 | 3226 | 3119 | 3010 | 2888 | 2756 | 2710 | 2724 | 2739 | 2753 |
| mu0=1.1 | 3717 | 3650 | 3581 | 3508 | 3430 | 3349 | 3262 | 3171 | 3074 | 2977 | 2867 | 2750 | 2727 | 2742 | 2756 | 2771 |

b) $L_{OV}$ (µm)

| | Vt=5.9 V | Vt=5.92 V | Vt=5.94 V | Vt=5.96 V | Vt=5.98 V | Vt=6 V | Vt=6.02 V | Vt=6.04 V | Vt=6.06 V | Vt=6.08 V | Vt=6.1 V | Vt=6.12 V | Vt=6.14 V | Vt=6.16 V | Vt=6.18 V | Vt=6.2 V |
|---|---|---|---|---|---|---|---|---|---|---|---|---|---|---|---|---|
| mu0=0.94 | 13.1699 | 14.6752 | 16.9795 | 20.8868 | 28.5457 | NA | NA | NA | NA | NA | NA | NA | NA | NA | NA | NA |
| mu0=0.96 | 4.8979 | 4.7725 | 4.6397 | 4.4983 | 4.3453 | 4.1732 | NA | 3.7606 | 3.4874 | 3.1401 | 2.7492 | NA | NA | NA | NA | NA |
| mu0=0.98 | 3.0905 | 2.9158 | 2.7331 | 2.5414 | 2.3392 | 2.1243 | 1.8955 | 1.6509 | 1.3868 | 1.0996 | 0.8078 | 0.4627 | 0.0840 | 0.0349 | 0.0200 | 0.0051 |
| mu0=1 | 2.3864 | 2.2271 | 2.0638 | 1.8960 | 1.7232 | 1.5450 | 1.3609 | 1.1704 | 0.9729 | 0.7678 | 0.5660 | 0.3436 | 0.1265 | 0.1111 | 0.0957 | 0.0804 |
| mu0=1.02 | 2.0464 | 1.9037 | 1.7589 | 1.6119 | 1.4625 | 1.3106 | 1.1560 | 0.9985 | 0.8380 | 0.6744 | 0.5153 | 0.3443 | 0.2038 | 0.1880 | 0.1722 | 0.1564 |
| mu0=1.04 | 1.8680 | 1.7383 | 1.6076 | 1.4758 | 1.3428 | 1.2087 | 1.0732 | 0.9365 | 0.7984 | 0.6609 | 0.5240 | 0.3812 | 0.2819 | 0.2656 | 0.2493 | 0.2332 |
| mu0=1.06 | 1.7752 | 1.6552 | 1.5349 | 1.4141 | 1.2928 | 1.1711 | 1.0488 | 0.9260 | 0.8026 | 0.6821 | 0.5592 | 0.4338 | 0.3607 | 0.3439 | 0.3272 | 0.3106 |
| mu0=1.08 | 1.7328 | 1.6203 | 1.5078 | 1.3952 | 1.2825 | 1.1697 | 1.0568 | 0.9438 | 0.8307 | 0.7216 | 0.6087 | 0.4950 | 0.4401 | 0.4229 | 0.4058 | 0.3887 |
| mu0=1.1 | 1.7229 | 1.6162 | 1.5096 | 1.4032 | 1.2969 | 1.1908 | 1.0849 | 0.9791 | 0.8735 | 0.7722 | 0.6668 | 0.5614 | 0.5202 | 0.5026 | 0.4850 | 0.4675 |

c) γ

| | Vt=5.9 V | Vt=5.92 V | Vt=5.94 V | Vt=5.96 V | Vt=5.98 V | Vt=6 V | Vt=6.02 V | Vt=6.04 V | Vt=6.06 V | Vt=6.08 V | Vt=6.1 V | Vt=6.12 V | Vt=6.14 V | Vt=6.16 V | Vt=6.18 V | Vt=6.2 V |
|---|---|---|---|---|---|---|---|---|---|---|---|---|---|---|---|---|
| mu0=0.94 | 6.8e-01 | 7.1e-01 | 7.6e-01 | 8.2e-01 | 9.2e-01 | NA | NA | NA | NA | NA | NA | NA | NA | NA | NA | NA |
| mu0=0.96 | 4.0e-01 | 4.0e-01 | 4.0e-01 | 3.9e-01 | 3.9e-01 | 3.8e-01 | NA | 3.6e-01 | 3.5e-01 | 3.3e-01 | 3.0e-01 | NA | NA | NA | NA | NA |
| mu0=0.98 | 2.9e-01 | 2.8e-01 | 2.7e-01 | 2.6e-01 | 2.5e-01 | 2.3e-01 | 2.1e-01 | 1.9e-01 | 1.7e-01 | 1.4e-01 | 1.1e-01 | 6.4e-02 | 6.1e-03 | 9.9e-05 | 9.9e-05 | 9.9e-05 |
| mu0=1 | 2.3e-01 | 2.2e-01 | 2.1e-01 | 1.9e-01 | 1.8e-01 | 1.7e-01 | 1.5e-01 | 1.3e-01 | 1.1e-01 | 8.8e-02 | 6.3e-02 | 3.3e-02 | 9.9e-05 | 9.9e-05 | 9.9e-05 | 9.9e-05 |
| mu0=1.02 | 1.9e-01 | 1.8e-01 | 1.7e-01 | 1.6e-01 | 1.4e-01 | 1.3e-01 | 1.1e-01 | 9.9e-02 | 8.2e-02 | 6.3e-02 | 4.3e-02 | 2.0e-02 | 9.9e-05 | 9.9e-05 | 9.9e-05 | 9.9e-05 |
| mu0=1.04 | 1.6e-01 | 1.5e-01 | 1.4e-01 | 1.3e-01 | 1.2e-01 | 1.1e-01 | 9.3e-02 | 7.9e-02 | 6.5e-02 | 4.9e-02 | 3.2e-02 | 1.3e-02 | 9.9e-05 | 9.9e-05 | 9.9e-05 | 9.9e-05 |
| mu0=1.06 | 1.4e-01 | 1.3e-01 | 1.2e-01 | 1.1e-01 | 1.0e-01 | 9.0e-02 | 7.9e-02 | 6.6e-02 | 5.3e-02 | 3.9e-02 | 2.5e-02 | 8.8e-03 | 9.9e-05 | 9.9e-05 | 9.9e-05 | 9.9e-05 |
| mu0=1.08 | 1.3e-01 | 1.2e-01 | 1.1e-01 | 9.9e-02 | 8.9e-02 | 7.9e-02 | 6.8e-02 | 5.7e-02 | 4.5e-02 | 3.3e-02 | 2.0e-02 | 5.8e-03 | 9.9e-05 | 9.9e-05 | 9.9e-05 | 9.9e-05 |
| mu0=1.1 | 1.1e-01 | 1.1e-01 | 9.7e-02 | 8.8e-02 | 7.9e-02 | 7.0e-02 | 6.0e-02 | 4.9e-02 | 3.9e-02 | 2.8e-02 | 1.6e-02 | 3.6e-03 | 9.9e-05 | 9.9e-05 | 9.9e-05 | 9.9e-05 |

In order to evaluate the goodness of the fitting resulting from the algorithm outlined above, we define as a figure of merit the quantity *err*, with the aim of weighting the goodness of fitting for both the current and the contact resistance:

1. We calculate the quantity $err_I = \sum_{V_g=23V}^{40V} \left( \frac{I - I_{fitted}}{I_{fitted}} \right)^2$

2. We calculate the quantity $err_{R_c} = \sum_{V_g=23V}^{40V} \left( \frac{R_c - R_{c,fitted}}{R_{c,fitted}} \right)^2$

3. We define $err = err_{R_c} + err_I$

The set of calculated quantities *err* for each combination of parameters $\mu_0$ and $V_t$ is presented In Table S3, where the acceptable values are highlighted in red with the same criterion as Table S2 above.

**Table S3:** Calculated values for *err* according to our algorithm. Values corresponding to the combinations where $L_{ov}$ is within the acceptable range (according to a second independent measurement) are highlighted in red.

| err | Vt=5.9 V | Vt=5.92 V | Vt=5.94 V | Vt=5.96 V | Vt=5.98 V | Vt=6 V | Vt=6.02 V | Vt=6.04 V | Vt=6.06 V | Vt=6.08 V | Vt=6.1 V | Vt=6.12 V | Vt=6.14 V | Vt=6.16 V | Vt=6.18 V | Vt=6.2 V |
|---|---|---|---|---|---|---|---|---|---|---|---|---|---|---|---|---|
| mu0=0.94 | 0.05238 | 0.06414 | 0.08314 | 0.11798 | 0.19421 | NA | NA | NA | NA | NA | NA | NA | NA | NA | NA | NA |
| mu0=0.96 | 0.00845 | 0.00833 | 0.00819 | 0.00801 | 0.00780 | 0.00752 | NA | 0.00674 | 0.00615 | 0.00537 | 0.00451 | NA | NA | NA | NA | NA |
| mu0=0.98 | 0.00331 | 0.00312 | 0.00291 | 0.00269 | 0.00245 | 0.00221 | 0.00195 | 0.00169 | 0.00144 | 0.00120 | 0.00101 | 0.00091 | 0.00127 | 0.00139 | 0.00141 | 0.00143 |
| mu0=1 | 0.00188 | 0.00176 | 0.00163 | 0.00150 | 0.00138 | 0.00125 | 0.00113 | 0.00102 | 0.00093 | 0.00086 | 0.00084 | 0.00092 | 0.00124 | 0.00126 | 0.00127 | 0.00129 |
| mu0=1.02 | 0.00129 | 0.00122 | 0.00114 | 0.00107 | 0.00100 | 0.00093 | 0.00087 | 0.00082 | 0.00079 | 0.00078 | 0.00082 | 0.00093 | 0.00114 | 0.00115 | 0.00116 | 0.00118 |
| mu0=1.04 | 0.00100 | 0.00095 | 0.00091 | 0.00086 | 0.00082 | 0.00078 | 0.00076 | 0.00074 | 0.00074 | 0.00076 | 0.00081 | 0.00093 | 0.00106 | 0.00107 | 0.00108 | 0.00109 |
| mu0=1.06 | 0.00084 | 0.00081 | 0.00078 | 0.00075 | 0.00073 | 0.00071 | 0.00070 | 0.00070 | 0.00071 | 0.00074 | 0.00080 | 0.00091 | 0.00099 | 0.00099 | 0.00100 | 0.00101 |
| mu0=1.08 | 0.00074 | 0.00072 | 0.00070 | 0.00068 | 0.00067 | 0.00066 | 0.00066 | 0.00067 | 0.00069 | 0.00073 | 0.00079 | 0.00088 | 0.00093 | 0.00094 | 0.00094 | 0.00095 |
| mu0=1.1 | 0.00067 | 0.00066 | 0.00065 | 0.00064 | 0.00064 | 0.00064 | 0.00064 | 0.00066 | 0.00068 | 0.00072 | 0.00077 | 0.00085 | 0.00088 | 0.00088 | 0.00089 | 0.00090 |

The best fittings of the experimental data curves when combined with the constraints on the acceptable range of $L_{ov}$ are identified for $V_T$ = 6.1 V and 1.02 cm²/Vs < $\mu_0$ < 1.08 cm²/Vs (Figure S7): indeed, the range for $\mu_0$ ~ 1 cm²/Vs is consistent with independent reports for the adopted semiconducting polymer P(NDI2OD-T2)[10] and the range for $V_T$ is reasonable and consistent with the measured transfer curves for our devices. In addition we verified that the injection length $L_0$ is larger than $L_{ov}$, as needed for equation (2) to hold (indeed Equation (2) is a very good approximation of Equation (1) already starting from $L_{ov} = L_0$, where the relative error is as low as 1.54%).[11]

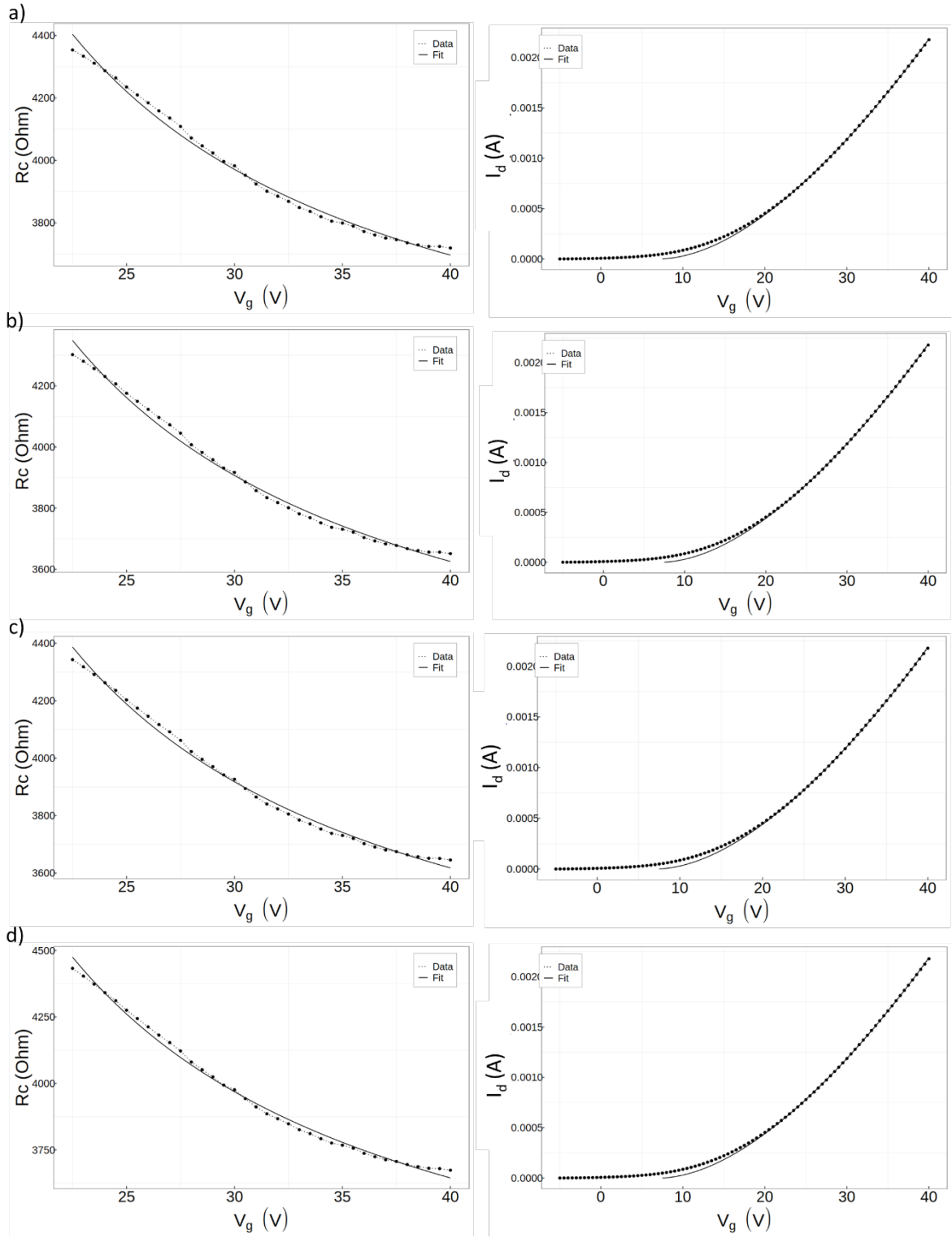

**Figure S7:** Experimental data and fitted curves as a result of our algorithm, for the combinations corresponding to $V_T = 6.1$ V and a) $\mu_0 = 1.02$ cm$^2$/Vs, b) $\mu_0 = 1.04$ cm$^2$/Vs, c) $\mu_0 = 1.06$ cm$^2$/Vs, d) $\mu_0 = 1.08$ cm$^2$/Vs.

For these ranges, 0.52 µm < $L_{ov}$ < 0.61 µm, γ is approximately zero (below 0.043) and 2888 Ω < $R_{c,const}$ < 3072 Ω. Different choices for gamma returned extremely similar results:

- For $\gamma = \frac{1}{4}\tilde{\gamma}$: 0.54 µm < $L_{ov}$ < 0.61 µm and 2912 Ω < $R_{c,const}$ < 3022 Ω,

- For $\gamma = \frac{3}{4}\tilde{\gamma}$: 0.41 µm < $L_{ov}$ < 0.57 µm and 2768 Ω < $R_{c,const}$ < 3126 Ω.

In conclusion, we estimate for our high-frequency OFETs an $R_c$ ~ 3600-3700 Ω at a bias voltage of 40 V. Such $R_c$ is composed of a constant component estimated as $R_{c,const}$ ~ 3000 Ω and of a bias-dependent component calculated through Equation 2. The corresponding width-normalized contact resistance for our OFET is thus $R_cW$ ~ 300 Ωcm at a bias of 40 V in the saturation regime.

**Consistence of $R_cW$ with the theoretical predictions for $f_t$**

The experimental values reported here for $f_t$ can be analyzed in the frame of a recently reported theoretical roadmap for high-frequency operation of organics.[12] With the model of that work, we express:

$$f_t = \frac{\mu_{eff}(V_G - V_T)}{2\pi L \left(\frac{2}{3}L + 2L_{ov}\right)}$$

where the parameters are defined analogously to the definitions in the main text, and

$$\mu_{eff} = \frac{\mu_0}{1 + \frac{\mu_0 R_c W}{L} C_{ins}(V_G - V_T)}$$

The contact resistance is described in accordance with the current-crowding model as in Equation (2), and considered as fully insisting on the source electrode.

When plugging in the parameters of the transistors of this work, as determined by the method described in the previous section, we obtain $f_t \sim$ 138 - 146 MHz, which is consistent with the experimental measurement. We remark that, in the adopted model, the voltage dependence of the mobility on the gate voltage is not accounted for. However, such contribution is effective only at a second order, since $\gamma < 0.043$.